\documentclass[article,floatfix,twocolumn]{revtex4}
\usepackage{graphicx}
\usepackage{color}
\usepackage{amsmath}
\newcommand{\ket}[1]{\left| #1 \right\rangle}

\newcommand{\Tr}{\mathrm{Tr}}
\newcommand{\Nmax}{N}
\newcommand{\Nrand}{N_{{\rm rand}}}
\newcommand{\Prob}{\mathrm {Prob}}
\newcommand{\rin}{\rho^{{\rm in}}}

\newcommand{\expect}[1]{\left\langle#1\right\rangle}

\begin{document}
\title{The power of random measurements: measuring $\Tr\rho^n$ on single copies of $\rho$}
\author{S.J.~van Enk$^{1,2}$ and C.W.J.~Beenakker$^3$}
\affiliation{$^1$Department of Physics and
Oregon Center for Optics\\
University of Oregon, Eugene OR 97403\\
$^2$ Institute for Quantum Information, California Institute of Technology\\
 $^3$Instituut-Lorentz, Universiteit Leiden, P.O. Box 9506, 2300 RA Leiden, The Netherlands }
\begin{abstract}
While it is known that $\Tr \rho^n$ can be measured directly (i.e., without first reconstructing the density matrix) by performing joint measurements on $n$ copies
of the same state $\rho$, it is shown here that random measurements on single copies suffice, too.
Averaging over the random measurements directly yields  estimates of $\Tr \rho^n$, even when it is not known what measurements were actually performed (so that $\rho$ cannot be reconstructed). 
\end{abstract}
\maketitle
The standard textbook quantum measurement of an observable $\hat{O}$ on a given quantum system
produces an estimate of the expectation value $\Tr(\rho\hat{O})$, where $\rho$ is the density matrix of the system.
 This expectation value is linear in $\rho$.
As is well-known by now \cite{Horo2002,Horo2003,Carteret,Brun2004,Mintert2007}, {\em non}linear functions of the density matrix $\rho$, such as the purity $p_2=\Tr\rho^2$ and its cousins
$p_n=\Tr\rho^n$ for $n>2$, can be measured directly, too, without first having to reconstruct the whole density matrix.  For this direct measurement method to work one needs $n$ quantum systems that are all in the same state $\rho$,  plus the ability to perform the appropriate joint measurement(s) on those multiple copies.
 
Here we point out that estimates of the same nonlinear quantities can be obtained from {\em random} measurements on {\em single} copies as well. A random measurement can be assumed to be
 implemented by performing a random unitary rotation on the single copy (possibly including an ancilla which starts off in a standard state), followed by a fixed measurement on the single copy (and possibly on the ancilla).
By averaging the measurement results over the random unitaries, one can directly infer estimates of $\Tr \rho^n$ [with $n=2\ldots M$, with $M$ the Hilbert space dimension of the system of interest], without having to reconstruct the density matrix. One point of the averaging procedure is that one does not have to know which random unitaries were in fact applied, and as a consequence one cannot reconstruct the density matrix in that case.
An example of a random measurement is furnished by  intensity measurements of speckle patterns resulting from light (be it two photons, or a single photon, or a coherent laser beam) propagating through a disordered medium \cite{Beenakker2009,Peeters2011}, and in that case the purity $p_2$ can (and was indeed) inferred directly from those measurements (see also \footnote{A different case is the experiment of M.~Munroe {\em et al.}, Phys. Rev. A {\bf 52}, R924 (1995) where diagonal density matrix elements  in the photon-number basis (hard to measure directly) were obtained by phase-averaging more straightforward quadrature  measurements.}).

There is an important difference between the known direct method and the current random method in what quantity exactly is estimated. Suppose one's source does not produce the same state every single time, but instead a state $\rho_j$ at try $j$.
In this case standard quantum measurements of a given observable on $J$ instances $j=1\ldots J$ can still be described by a single density matrix, namely, the mean $\bar{\rho}=\sum_j \rho_j/J$. Since the random method only involves measurements on single copies, it produces, likewise, an estimate of $\Tr(\overline{\rho}^n)$. This requires no assumption about the quantum systems being uncorrelated or unentangled with each other, since $\rho_j$ is obtained by tracing out all degrees of freedom except those of system $j$.

On the other hand, a direct measurement  would yield an estimate of
$\Tr(\hat{S}\overline{\rho_{j,j+1\ldots, j+n-1}})$ instead, where
$\rho_{j,j+1\ldots j+n-1}$ is the joint density matrix of $n$ systems $j,j+1,\ldots, j+n-1$, and
$\hat{S}$ is the cyclical shift operator, which acts on the basis states of the $n$ quantum 
systems as $\hat{S}|\psi_j\rangle|\psi_{j+1}\rangle\ldots|\psi_{j+n-1}\rangle=
|\psi_{j+1}\rangle|\psi_{j+2}\rangle\ldots|\psi_{j}\rangle.$
It is only under the assumption that the states of the $n$ systems are identical and independent ({\em i.i.}) that the direct measurement yields 
$\Tr\rho^n$. In fact, the direct measurement is
eminently suited for detecting that
the states are {\em not} identical \cite{Lucia}.
Although the assumption of {\em i.i.} states is standard, it is only recently that 
precise conditions have been stated under which the approximate
{\em i.i.} character can be inferred \cite{Renner2007}. The required
permutation invariance is easily enforced when performing measurements on single copies, but not when performing joint measurements on multiple copies \cite{vanEnk2009}. 
Avoiding this difficulty is the main advantage of the random method.

An $\Nmax\times \Nmax$ random unitary matrix, distributed according to the Haar measure, can be easily constructed by the method presented in \cite{Random}. One first constructs
a matrix whose elements are independent complex Gaussian variables, and one then performs an orthogonalization of the resulting random matrix (where one small pitfall needs to be avoided \cite{Random}).
We first consider approximate results for random unitaries, because the resulting expressions are quite simple, and subsequently we will give the more involved exact results.

If we consider an arbitrary submatrix $V$ (of size $M$) of $U$ (of size $N$), with $M\ll N$  \cite{Zycz2000}, then the real and imaginary parts of its matrix elements can still be very well approximated  by independent and normally distributed numbers if $N$ is large. With this Gaussian approximation we can compute the following {\em averages} (we indicate averages over the distribution of random unitaries by $\expect{.}$): first, we have 
\begin{eqnarray}
\expect{V_{kl}V^*_{mn}}=
\delta_{km}\delta_{ln}/\Nmax.
\end{eqnarray}
Here and in all of the following we assume we have picked some basis $\{|k\rangle\}$, and we write all matrix elements w.r.t.\ that basis.
The normalization factor $1/\Nmax$ follows immediately from the fact that $U$, of which $V$ is a submatrix, is unitary, so that
$\sum_{l=1}^{\Nmax}U_{kl}U^*_{ml}=\delta_{km}$.
Higher-order averages follow from the Isserlis (``Gaussian-moments'') theorem \cite{1918}. In particular, the only nonzero averages arise from products of $2K$ factors of the form
\begin{eqnarray}
\expect{V_{k_1l_1}\ldots V_{k_K l_K}V^*_{m_1n_1}\ldots V^*_{m_K n_K}}=
\nonumber\\\frac{
\sum_{{\rm all\,pairs}(i,j)} \delta_{k_i m_j}\delta_{l_i n_j}}{\Nmax^K}.\label{Uaver}
\end{eqnarray}
We now apply the preceding approximate results to the following scenario.
Consider an ``input'' density matrix $\rin$ of size
$M\times M$. 
Embed the system in a larger Hilbert space of size $\Nmax$, by constructing a new $\Nmax\times\Nmax$ density matrix by adding zero matrix elements. 
 Then apply a random unitary $U$ to the larger matrix.
Finally, consider measurements in a fixed $M$-dimensional (sub)basis $\{\ket{k}\}$. The probability $\Prob(k)$ of finding measurement outcome $k$ is given by 
 \begin{equation}
\Prob(k)=\sum_{m,n}\rin_{mn}V_{mk} V^*_{nk}. 
\end{equation}
This expectation value depends on what $V$ is, of course, but its average is given simply by
 \begin{equation}
\expect{\Prob(k)}=\sum_{m,n}\rin_{mn}\expect{V_{mk} V^*_{nk}}=1/\Nmax, 
\end{equation}
where we used that $\Tr(\rin)=\sum_m\rin_{mm}=1$.
Defining $P_n(k)=\expect{\Prob(k)^n}$, the following averages are obtained by using the Isserlis theorem (up to order $n=4$; subsequent orders can be easily obtained, too, but for our purposes this will do)
\begin{subequations}\label{pk}
\begin{eqnarray}
P_2(k)&=&\left[1+p_2\right]/\Nmax^2,\\
P_3(k)&=&\left[1+3p_2+2p_3\right]/\Nmax^3,\\
P_4(k)&=&\left[1+3p_2^2+6p_2+8p_3+6p_4\right]/\Nmax^4,
\end{eqnarray}
\end{subequations}
where we defined $p_n=\Tr\left((\rin)^n\right)$.
Inverting these equations gives estimates of $p_n$ in terms of the measurable quantities on the left-hand sides. We denote those estimates by an overbar, e.g., $\bar{p}_2=\Nmax^2P_2(k)-1$.
We refrain from giving the other inverse relations now, as we will give the {\em exact} relations below in (\ref{pkall}).

We can also compute standard deviations in the (mean) estimates. For example, assuming
we average the results for one value of $k$ over $\Nrand$ random unitaries, then the statistical error in the estimate of the purity is 
\begin{equation}
\sqrt{\Nrand-1}\Delta(\bar{p}_2)=
\sqrt{4p_2+2p_2^2+8p_3+6p_4}.\label{Delta}
\end{equation}
This is an increasing function of $p_2, p_3, p_4$, so that the variance is largest for a pure state and smallest
for the totally mixed state $\rin=\openone/M$. 

In an actual experiment one may not know exactly what the values of $\Nmax$ and/or $M$ are (for instance, this is the case in the speckle experiments of Refs.~\cite{Beenakker2009,Peeters2011}). In such a case $\Nmax$ can be directly estimated from
$P_1(k)$ through $\Nmax= 1/P_1(k)$. So, we would use
\begin{eqnarray}
\tilde{p}_2&=&\frac{P_2(k)}{P_1(k)^2}-1,\label{pknum}
\end{eqnarray}
 instead (such estimates we indicate by a tilde). Now this estimate $\tilde{p}_2$ has a {\em smaller} variance than $\bar{p}_2$ has, simply because the errors in $P_1(k)$ and $P_2(k)$ are positively correlated. It is, therefore, better to use $\tilde{p}_2$ as estimate for $p_2$, even when $\Nmax$ is in principle known. The numerical results given below will confirm this, also for the exact result for $\tilde{p}_2$. For the estimates $\tilde{p}_3$ and $\tilde{p}_4$, however, there is not much difference between the two methods. 
 
When $\Nmax$ is not very large,  equations (\ref{Uaver}) and hence (\ref{pk}) are not correct. The exact results, which can be extracted from Refs.~\cite{Collins} and \cite{Puchala}, are still given by (\ref{pk}) upon multiplication of $P_n(k)$ by the correction factor $C_n$, where
\begin{eqnarray}
C_n&=&(1+1/\Nmax)(1+2/\Nmax)\ldots(1+(n-1)/\Nmax).\label{corr}
\end{eqnarray}
Note that these factors depend only on $\Nmax$, not on $M$, and the results are valid even when $M=N$.
This then leads to the inverse formulas:
\begin{subequations}\label{pkall}
\begin{eqnarray}
\bar{p}_2&=&D_2 P_2(k)-1,\\
\bar{p}_3&=&\tfrac{1}{2}\bigl[D_3 P_3(k)-1-3p_2\bigr],\\
\bar{p}_4&=&\tfrac{1}{6}\bigl[D_4P_4(k)-1-3p_2^2-6p_2-8p_3\bigr],
\end{eqnarray}
\end{subequations}
with $D_n=(N+n-1)!/(N-1)!$
Taking into account the correction factors (\ref{corr}) leads to different values for the statistical errors in estimates. It is still true that pure states lead to the largest errors; for those  we get 
\begin{equation}
\sqrt{\Nrand-1}\Delta(\bar{p}_2)=
\sqrt{\frac{24(1+1/N)}{(1+2/N)(1+3/N)}-4}.\label{Deltap}
\end{equation}
The right-hand side (slowly) increases with increasing $N$, from $\sqrt{52/7}$ for $N=4$ to $\sqrt{20}$ for $N\rightarrow\infty$.

In order to illustrate the method and the meanings of $\Nmax$ and $M$, we consider the following examples here:
{\bf (i)} Suppose we have a single photon occupying one of $M$ input modes. We then apply a random linear optics transformation that involves $\Nmax-M$ ancilla modes. The photon now ends up being coherently distributed over $\Nmax$ output modes. We then estimate the probability $\Prob(k)$ with which the photon ends up in one of a fixed set of $M$ output modes $k=1\ldots M$. This is
an example akin to that considered in \cite{Beenakker2009,Peeters2011}. 

{\bf (ii)} Suppose our system of interest consists of 2 qubits, so that $M=4$. Suppose we have an ancilla qubit
in a fixed state $|0\rangle$, and we apply a random unitary operation to the 3 qubits. In this case, $\Nmax=8$. We then perform measurements on each of the three qubits separately in the standard basis. We
measure the probability $\Prob(k)$ of the two qubits ending up in one of the $M=4$ combinations $k=00, 01, 10, 11$ {\em and} the ancilla ending up in $|0\rangle$ (thus measuring only a $M$-dimensional subspace). 

{\bf (ii')} There is no need for any ancillas if dealing with a fixed and known number of qubits, say $Q$. In that case, we simply have $\Nmax=M=2^Q$.
 We consider only case (ii') in the following numerical results.
 
We assume that we run an experiment with a fixed random (``unknown'') unitary of size $\Nmax$ sufficiently many times that we get a very good estimate of 
$\Prob(k)$ for each $k$ for the given unitary and the given input state (of size $M$). 
Subsequently we average  over $\Nrand$ random unitaries to obtain $P_n(k)=\expect{\Prob(k)^n}$. From those results we estimate the values of $p_2, p_3, p_4$.
 \begin{figure}[h]
  \begin{center}
    \includegraphics[width=191pt]{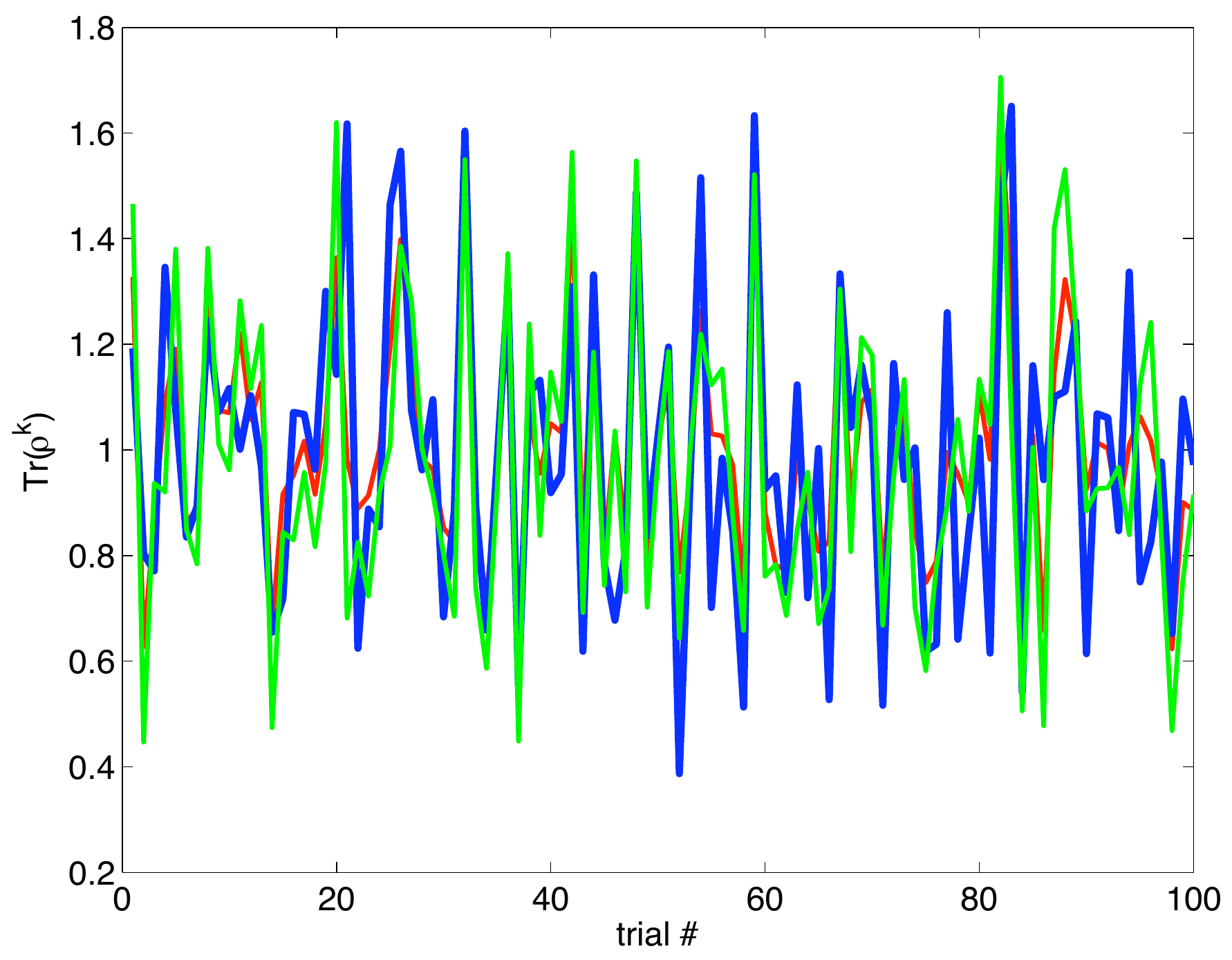}
  \end{center}
  \caption{This plot shows, for a pure two-qubit state, the estimated values of $p_2, p_3, p_4$ (blue: $p_2$, red: $p_3$, green: $p_4$) for 100 trials, each trial using just one value of $k$, containing an average over $\Nrand=100$ random unitaries, and using $\Nmax=4$  in (\ref{pkall}).
  The mean standard deviations (over 100 trials) were $\Delta\bar{p}_2=0.282$ [note that this agrees with the result (\ref{Deltap}), since  $\sqrt{52/7/99}\approx 0.274$], $\Delta\bar{p}_3=0.21$,
  $\Delta\bar{p}_2=0.29$. The mean estimates obtained by pooling all data from the 100 trials for $\bar{p}_n$ are: $\bar{p}_2=0.990$, $\bar{p}_3=1.01$ and
 $\bar{p}_4=1.02$, which are all consistent with their mean standard deviations (10 times smaller than the $\Delta \bar{p}_n$ given above). }
  \label{fi1}
\end{figure}
 \begin{figure}[h]
  \begin{center}
    \includegraphics[width=191pt]{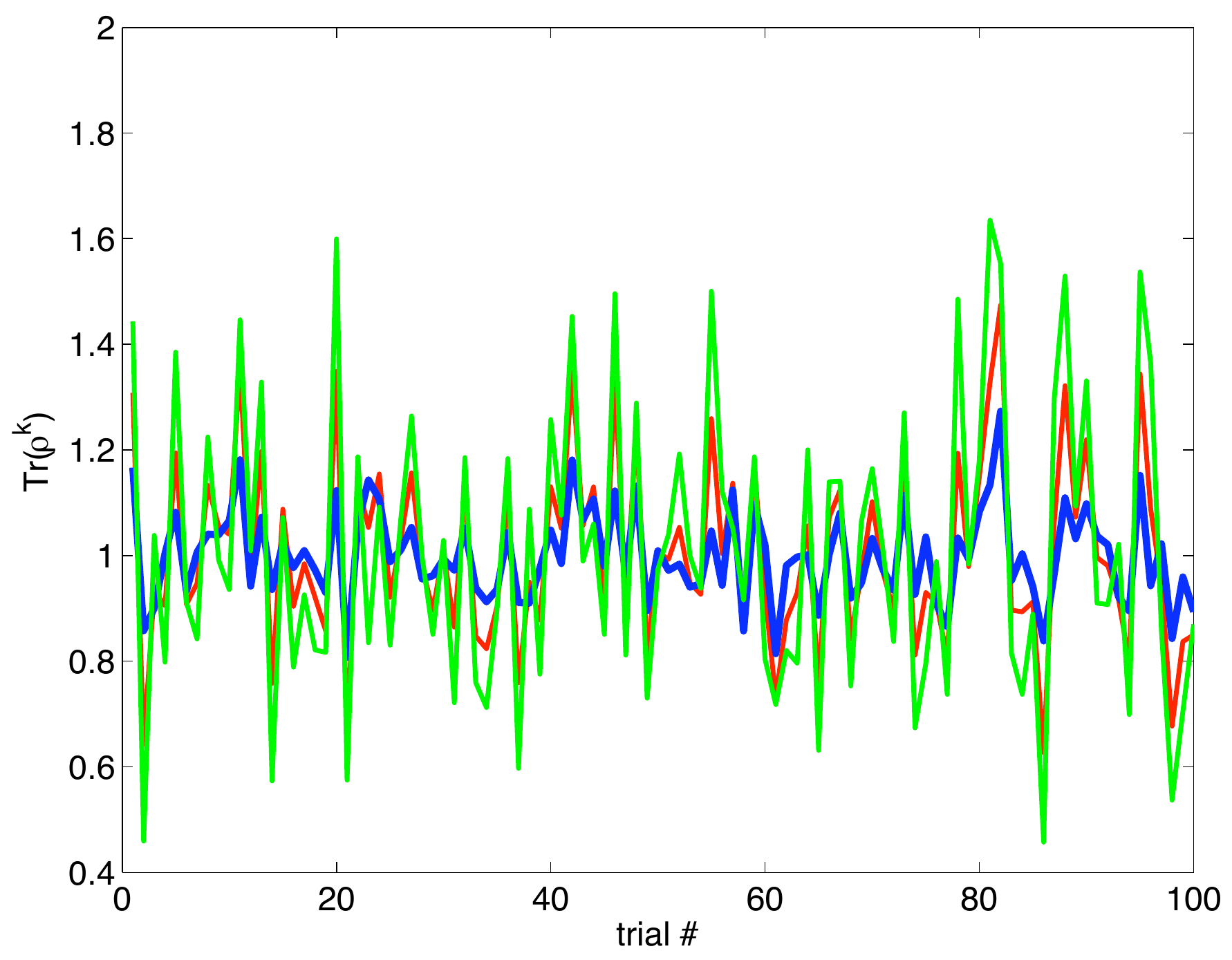}
  \end{center}
  \caption{Same as previous Figure, but using the estimate $\Nmax\approx 1/P_1(k)$ in (\ref{pkall}).
 Here we have  $\Delta\tilde{p}_2=0.09$, $\Delta\tilde{p}_3=0.17$,
  $\Delta\tilde{p}_4=0.27$. The mean estimates obtained from pooling all data (which are the {\em same} ``raw'' data as in Fig.~\ref{fi1}) from the 100 trials
  for $\tilde{p}_n$ are $\tilde{p}_2=1.005$ [which is indeed better than $\bar{p}_2$], $\tilde{p}_3=1.01$ and
 $\tilde{p}_4=1.02$, all consistent with the statistical errors in the mean (which are 10 times smaller than $\Delta\tilde{p}_n$).}
  \label{fi1b}
\end{figure}
The first example we consider corresponds to case (ii') mentioned above, where we have two qubits. In Figs.~\ref{fi1} and \ref{fi1b} we plot results for pure input states, where we use the results for just 1 value of $k$ to estimate $p_n$, in two different ways: using the exact value $\Nmax=4$ (Fig.~\ref{fi1}) or using the estimate $\Nmax\approx 1/\expect{\Prob(k)}$ (Fig.~\ref{fi1b}).  The results show how the latter method is more accurate for estimating purity.  The {\em same} data are used in the two Figures, so that all differences between them are entirely due to the different analysis of those data.
  This different analysis reduces the statistical variation in $\tilde{p}_2$, but not in $\tilde{p}_3$ and $\tilde{p}_4$.
In addition, the plots show that the statistical errors in $\tilde{p}_2, \tilde{p}_3, \tilde{p}_4$ are strongly  correlated in the latter case.

In the remaining figures we perform an additional average over the $M$ different values of $k$, leading to smaller (by a factor of about $\sqrt{M}$) error bars.

\begin{figure}[h]
  \begin{center}
    \includegraphics[width=191pt]{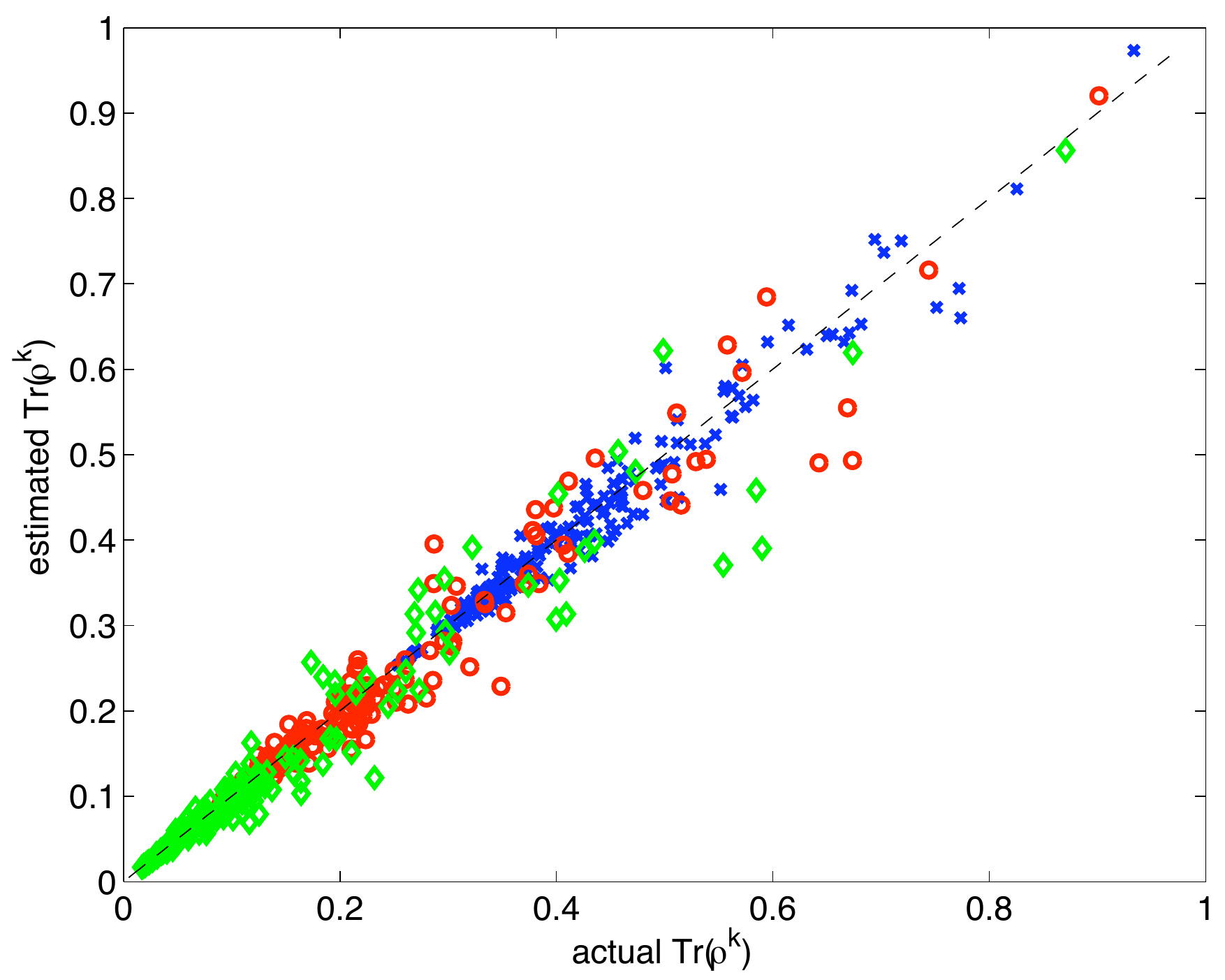}
  \end{center}
  \caption{Scatter plot of estimated values of $p_2$ (blue crosses), $p_3$ (red circles), and $p_4$ (green diamonds), versus their actual values for 
  200 randomly
  picked two-qubit input states. Here an average is taken over $\Nrand=30$ random unitaries, as well as over 4 measurement outcomes. For convenience, the dashed line gives the diagonal on which estimated and actual values agree.}
  \label{fi2}
\end{figure}

\begin{figure}[t]
    \includegraphics[width=191pt]{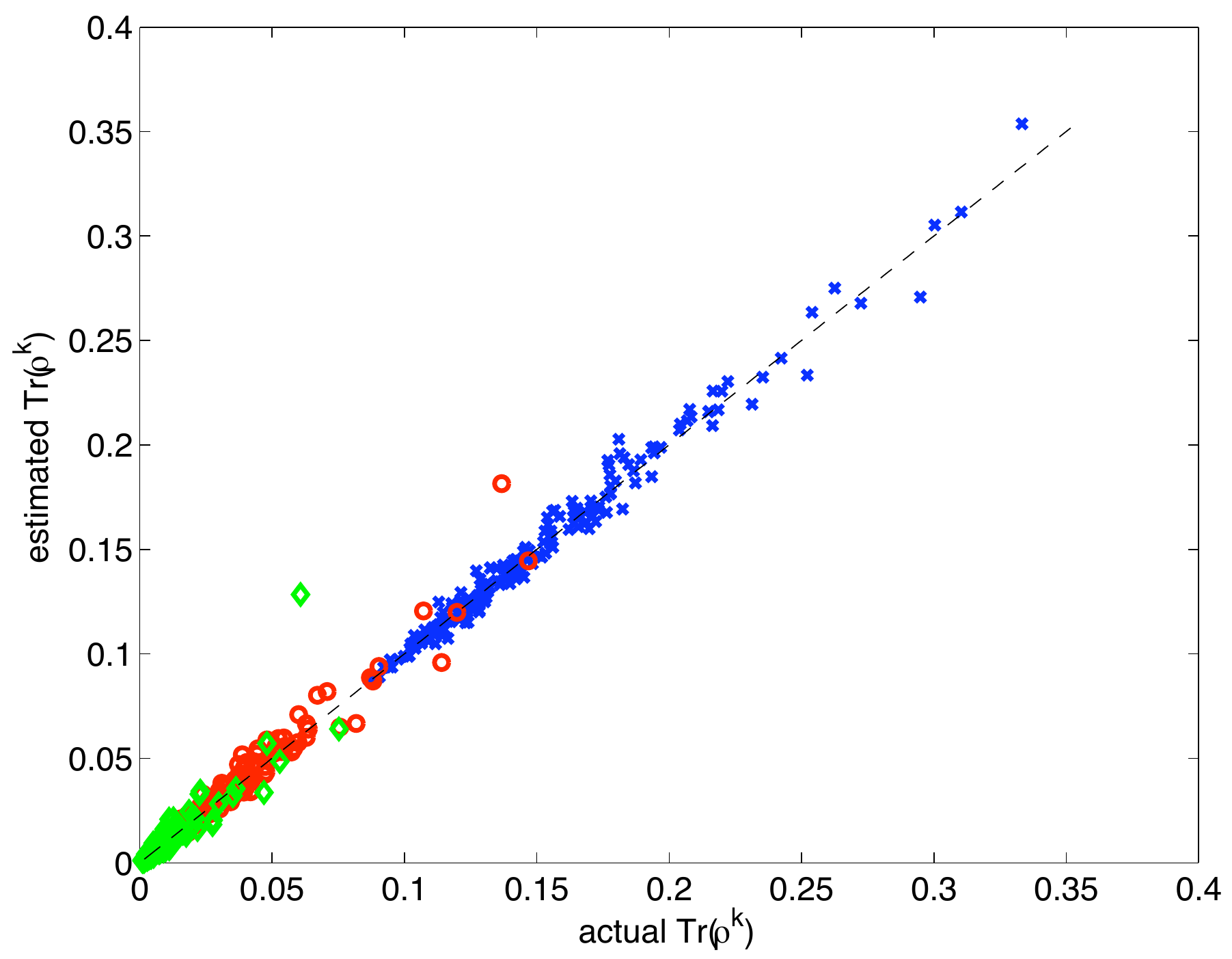}
  \caption{Same as Fig.~\ref{fi2}, but for randomly picked {\em five}-qubit states ($M=\Nmax=32$). Averaging over the same number of random unitaries (here $\Nrand=30$) produces a smaller statistical error for larger systems.}
  \label{fi3}
\end{figure}

Performing tomography on two qubits would require 15 independent (and known) measurements. Here we show that with just a moderate overhead one can  obtain good estimates of $p_2, p_3, p_4$ for {\em generic} (i.e.\ randomly picked \footnote{As only the eigenvalues of $\rho$ matter, the
  states were chosen
  according to a simple distribution, without any significance otherwise:
  first, $M$ uniformly distributed random numbers $(z_i)$ between 0 and 1 are picked;
  then $\rho$ is chosen to be the diagonal matrix diag$(z_i^E)/(\sum_i z_i^E)$, with $E$ chosen equal to 2 in Fig.~\ref{fi2} and $E=8$ in Fig.~\ref{fi3}. These choices were made so as to produce a wider spread of values for $\Tr\rho^n$ than do other more standard ensembles.}) states.
  In Fig.~\ref{fi2} results are displayed for 200 generic two-qubit states, using $\Nrand=30$. 

\begin{figure}[t]
    \includegraphics[width=191pt]{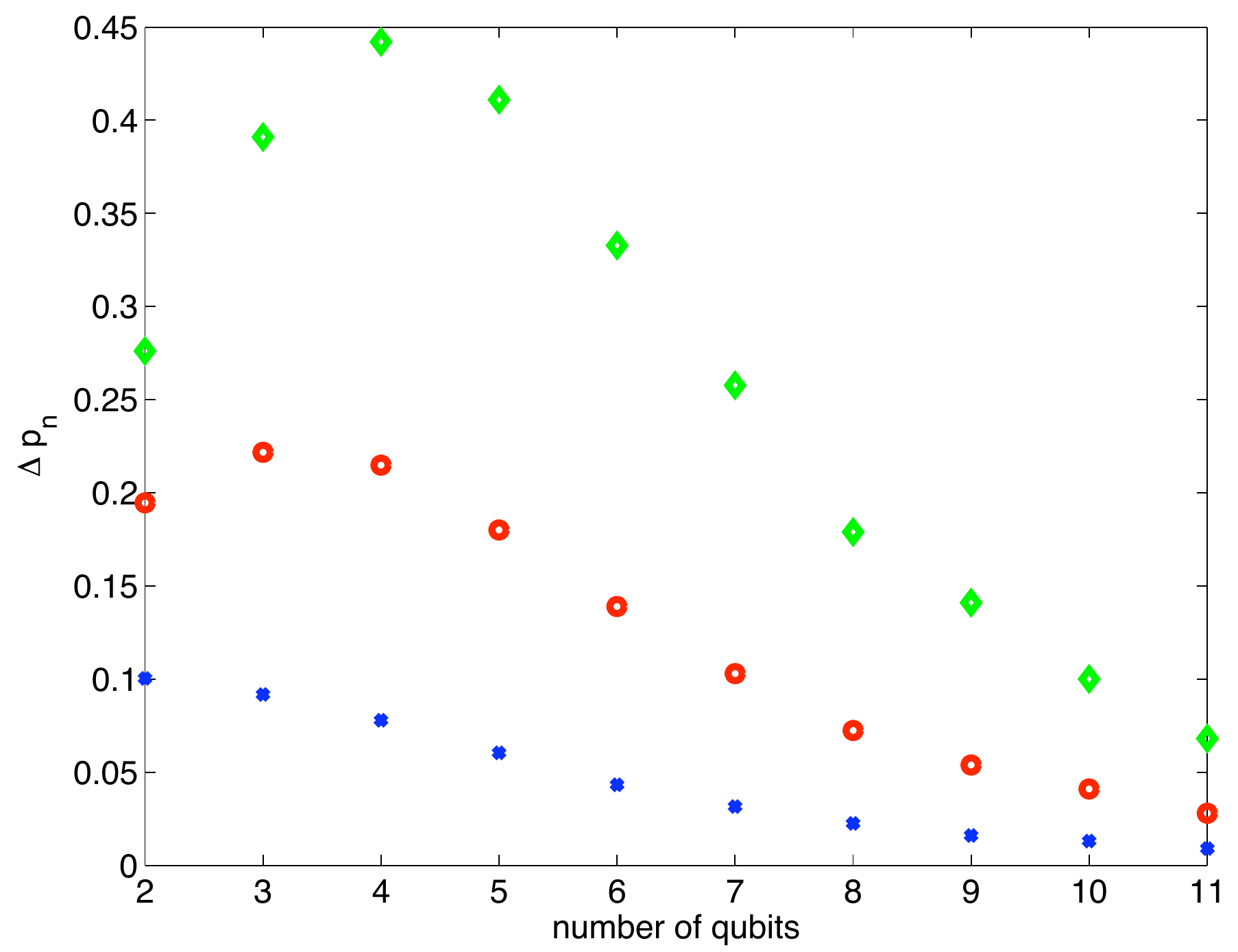}
  \caption{The standard deviation in the mean estimates for $\tilde{p}_2$ (blue crosses), $\tilde{p}_3$ (red circles), and $\tilde{p}_4$ (green diamonds) after averaging over $\Nrand=30$ random unitaries, for pure multi-qubit states, as a function of the number of qubits.}
  \label{fi5}
\end{figure}

In Fig.~\ref{fi3} we show (for five qubits) that the number of random unitaries needed to obtain a fixed-size error bar does not increase with the number of qubits. For $\Nrand=30$ one still obtains good estimates: in fact, the error bars {\em decrease} (roughly as $1/\sqrt{M}$) when going to more and more qubits, just because the number $M$ of measurement results one can average over increases exponentially with the number of qubits, while the variance (\ref{Deltap}) increases only very slowly. This is illustrated for pure multi-qubit states in Fig.~\ref{fi5}.
It shows that the statistical error in the estimate of $\Tr\rho^n$ for $n=2,3,4$ first increases with the number of qubits before (at $\geq n$ qubits) it starts to decrease monotonically.

In conclusion then, using the ideas of random matrix theory, we showed that
nonlinear functions of the density matrix such as $\Tr\rho^n$ can be directly obtained from appropriately averaged random  measurements on single copies. No assumptions are needed
on the independence of the copies, nor on their states being identical. This contrasts the random method with so-called direct measurements on $n$ identical copies \cite{Horo2002,Horo2003,Carteret,Brun2004,Mintert2007}. 

Moreover, one does not need to know which random measurements were actually performed, because the averaging procedure keeps all information about the eigenvalues 
of $\rho$, which is all that is needed to estimate $\Tr\rho^n$.
One does need to verify that the random unitaries have been drawn from the appropriate ensemble.
There are two tests one could perform: 
first of all, the definition of the ensemble is that it is unitarily invariant. This means in our context that all averages $\expect{\Prob(k)^n}$ should be independent of $k$. This is a statistically testable property. In addition, one can apply the random measurements to {\em known} input states, so that the values of those
$k$-independent averages are known.

Importantly, the number of unitaries over which one has to average in order to obtain a fixed error bar in the estimates of $\Tr\rho^n$ scales very favorably with the Hilbert space dimension of one's system: in fact, this number even tends to {\em decrease}. For two qubits this amounts to needing a small overhead as compared to full quantum-state tomography, but for larger systems (more than, say, four qubits) the random method requires (far) fewer resources than does full quantum-state tomography.

\bibliography{RandomU}

\end{document}